# High-pressure synthesis and superconductivity of $Ca_{2-x}Na_xCuO_2Cl_2$

Nikolai D. Zhigadlo*, Janusz Karpinski

*Laboratory for Solid StatePhysics, ETH Zurich, 8093 Zurich, Switzerland*

**Abstract**

$Ca_{2-x}Na_xCuO_2Cl_2$ is a structural analogue to the $La_{2-x}Sr_xCuO_4$ superconductor. By substituting $Na^+$ for $Ca^{2+}$ mobile holes are introduced into the $CuO_2$ planes. A series of experiments shows that the Na content $x$ depends not only on the pressure during synthesis but also on the synthesis temperature and time. $Ca_{2-x}Na_xCuO_2Cl_2$ crystals ($x\approx0.10$ and $T_{c,on}$=13.5 K) up to ~1x1x0.05 $mm^3$ have been grown by a flux method at 35 kbar and the temperature was slowly ramped from 1250ºC to 1050ºC within 8 h. Applying higher synthesis pressure (up to 55 kbar) and temperature (up to 1700ºC) results in polycrystalline samples with highest $T_{c,on}$ =28.0 K at $x\approx0.20$.

## 1. Introduction

$Ca_{2-x}Na_xCuO_2Cl_2$ is an ideal model system for studying of the electronic state of doped $CuO_2$ planes by surface-sensitive measurements such as ARPES, STM, STS because *(a) it is easy to cleave*, *(b) the hole concentration can be tuned by chemical means (resulting in either insulating or superconducting crystals)*, *(c) there is no orthorhombic distortion*, *(d) the structure is not modulated* [1,2]. Substituting Na for Ca is energetically so unfavorable that growth of the $Ca_{2-x}Na_xCuO_2Cl_2$ compounds is only possible under high pressure of several tens of kbar. Superconductivity in $Ca_{2-x}Na_xCuO_2Cl_2$ was first discovered by Hiroi *et al.* [3]. The first single crystals of $Ca_{2-x}Na_xCuO_2Cl_2$ with $x$=0.12 correspond to heavily underdoped superconductor have been grown by Kohsaka *et al.* [4]. The goal of our studies is to find optimal conditions for growing of $Ca_{2-x}Na_xCuO_2Cl_2$ crystals with a higher concentration of Na and with $T_c$ characteristic for optimally doped and overdoped regions.

* Corresponding author. Tel.:+41 1 633 2249; fax: +41 1 633 1072
*E-mail address*: zhigadlo@phys.ethz.ch (N.D.Zhigadlo)

## 2. Experimental

The non superconducting compound, $Ca_2CuO_2Cl_2$, was prepared in advance under ambient pressure. Powders of $Ca_2CuO_2Cl_2$, $NaClO_4$ (flux, Na source, and oxidizer) and NaCl (Na source) in a molar ratio of 1:0.2:0.2 were well mixed in a dry box and sealed in Au or Pt cylindrical capsules. High-pressure experiments have been performed in a cubic anvil and opposed anvil-type high-pressure devices at 35-55 kbar and 1250-1700ºC. High pressure products have been characterized by X-ray powder diffraction. Magnetic susceptibility data were collected by a SQUID magnetometer to check superconductivity of the high pressure samples.

## 3. Experimental results

$Ca_{2-x}Na_xCuO_2Cl_2$ crystals ($x\approx0.10$ and $T_{c,on}$=13.5 K) of size up to ~1x1x0.05 $mm^3$ have been grown by a flux method at 35 kbar and the temperature was slowly ramped from 1250ºC to 1050ºC within 8 h (Fig. 1). Small flux inclusions were often present in the crystals. Due to the extremely hydroscopic nature of the compound $Ca_{2-x}Na_xCuO_2Cl_2$, the crystals were handled in a dry box. The hydroscopic nature of samples means that the initial pristine surface layer is potentially more reactive and thus prone to surface aging from absorbates. The sample lifetime depends on vacuum conditions.

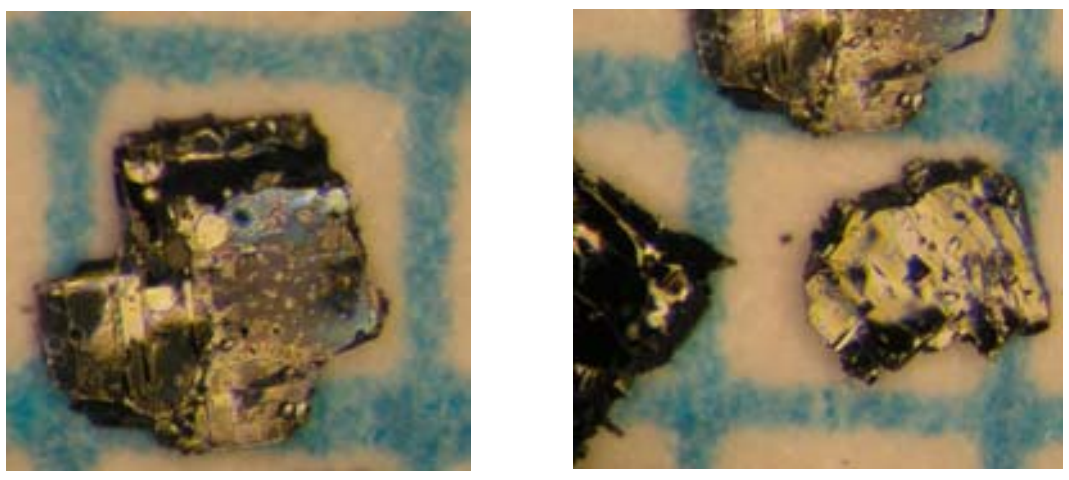

Fig. 1. A photographs of a $Ca_{2-x}Na_xCuO_2Cl_2$ crystals grown under high-pressure. The edge of the picture is about 1 mm.

A series of experiments at 45 kbar shows that the Na content depends not only on the synthesis pressure, as it was established before [3-5], but also on the reaction temperature and time (see vertical dash line in Fig. 2). For example, when the sample was compressed at 45 kbar and the temperature was slowly ramped from 1500ºC to 1300ºC within 5 h, the product shows $T_{c,on}$=28.0 K. Similar $T_c$ was observed when samples were synthesized at 55 kbar.

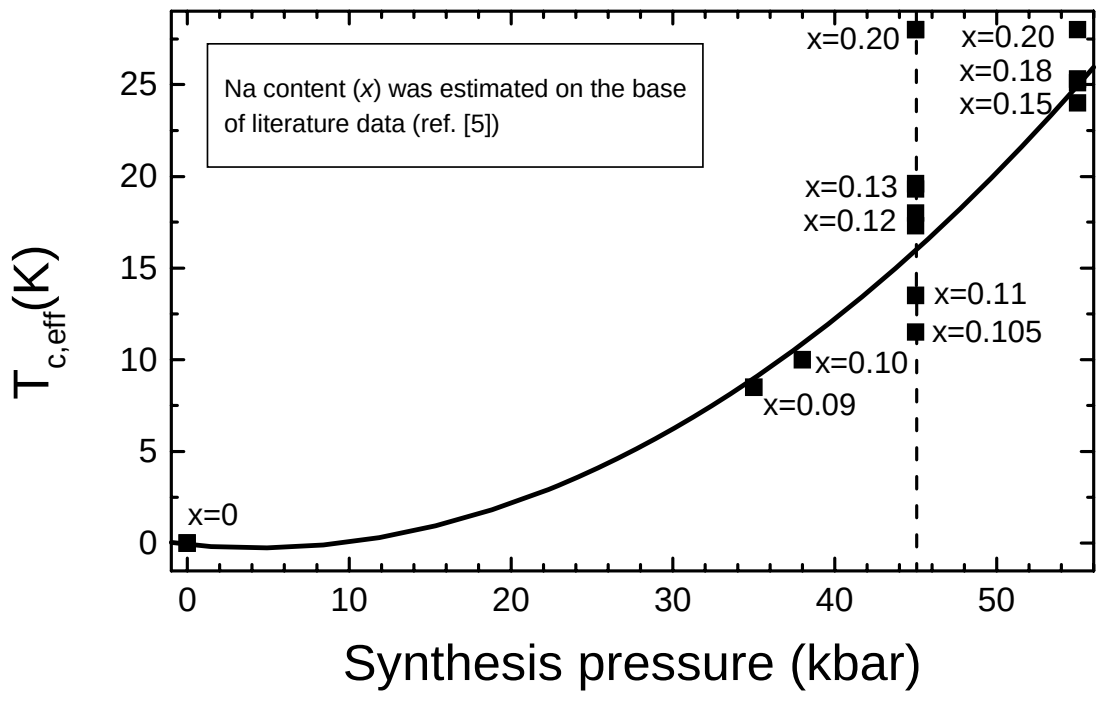

Fig. 2. The Na content ($x$) of $Ca_{2-x}Na_xCuO_2Cl_2$ was controlled by the synthesis pressure and temperature. As Na content increases $T_c$ changes systematically.

Fig. 3 shows the normalized diamagnetic signal for a series of $Ca_{2-x}Na_xCuO_2Cl_2$ samples with various Na content. $T_c$ varies systematically with the Na content. We note that the transition temperatures for crystals and polycrystalline samples originating from different parts of the capsule were almost identical. Compared to the non superconducting crystals, introducing holes and associated superconductivity is accompanied by an expansion of the unit cell along the $c$, and a contraction along the $a$ axis (Fig. 4). These features are characteristic of a $p$-type cuprate superconductor.

Further experiments are under way to optimize conditions to grow crystals with extended concentration of Na and $T_c$ characteristic for optimally doped and overdoped regions.

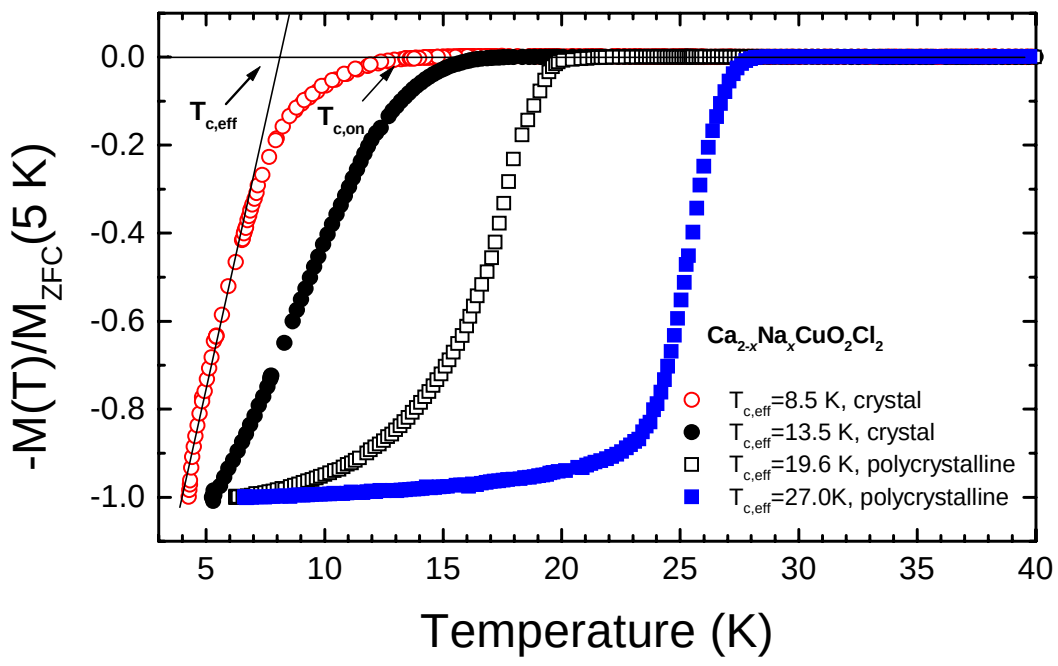

Fig. 3. Normalized diamagnetic signal for a series of $Ca_{2-x}Na_xCuO_2Cl_2$ samples with various Na contents grown at high-pressure.

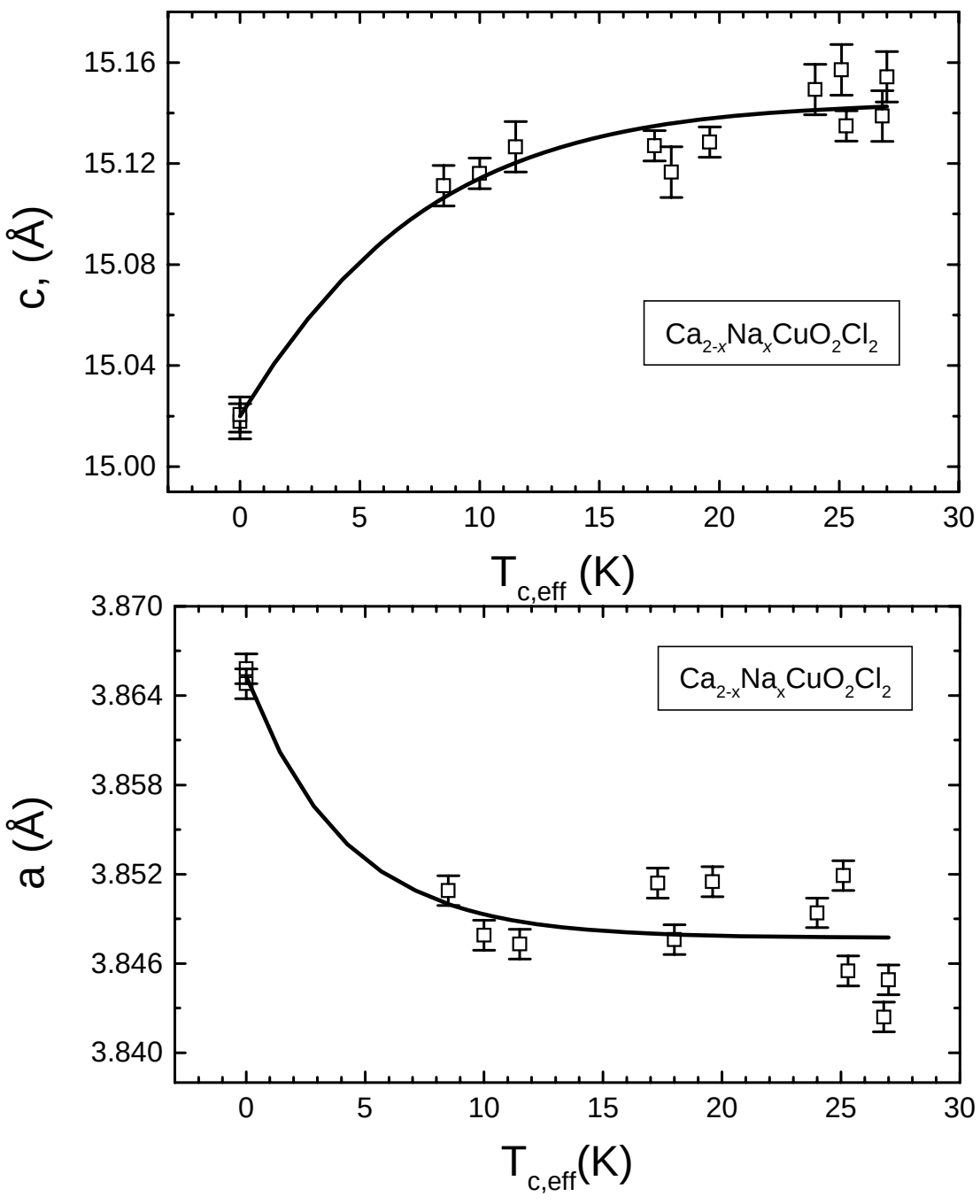

Fig. 4. The lattice constants versus $T_c$ for $Ca_{2-x}Na_xCuO_2Cl_2$. The $a$-axis shrank while the $c$-axis expanded (r($^{IX}Na^+$)=1.24Å; r($^{IX}Ca^{2+}$)=1.18Å